\documentstyle[aps,prl,epsf,twocolumn,epsfig,floats]{revtex}
\begin{document}
\draft


\wideabs{

\title{A family of sure-success quantum algorithms for solving a generalized
Grover search problem}
\author{Chia-Ren Hu}
\address{Department of Physics,
Texas A\& M University, College Station, Texas 77843-4242}
\date{Jan. 12, 2002}
\maketitle
\begin{abstract}
This work considers a generalization of Grover's search problem, {\it viz.}, to find any
one element in a set of acceptable choices which constitute a fraction $f$ of the total
number of choices in an unsorted data base. An infinite family of sure-success quantum
algorithms are introduced here to solve this problem, each member for a different
range of $f$.  The $n$th member of this family involves $n$ queries of the data base,
and so the lowest few members of this family should be very convenient algorithms
within their ranges of validity. The even member ${\cal A}_{2n}$ of the family covers
ever larger range of $f$ for larger $n$, which is expected to become the full range
$0\le f\le 1$ in the limit $n\rightarrow\infty$.

\end{abstract}
\pacs{PACS numbers: }
}  


Quantum computing as a new powerful approach to solve difficult computational problems
is still in its infancy. Only a handful useful algorithms have been proposed so far.
Most of them fall into two categories: Those for factorizing large integers,
and those for ``searching a needle in a haystack'', or finding the only acceptable
element in a large unsorted data base. The main idea in the former category is due to
Shor,~\cite{Shor} and in the latter category is due to Grover.~\cite{Grover} Here we
wish to consider a generalization of Grover's search problem, {\it viz.}, to find any
one element in a set of acceptable choices which form a fraction $f$ of the total number
of choices in an unsorted data base of size $N$.~\cite{note0} An infinite family
$\{{\cal A}_n\,|\,n = 1, 2, 3, \cdots\}$ of qunatum algorithms is introduced here, each
similar to one stage of Grover's algorithm, except that, unlike Grover's original
algorithm, which requires iteration to some optimal stage, which in general is still not
a sure-success algorithm,~\cite{note1} here each member is an independent sure-success
algorithm within its range of validity.  Each member of the family introduced here is
characterized by an iteration number, $n$, in the sense introduced in the original Grover
algorithm. This number is also the number of times the data base is queried.  Here we
only analyze four members of this family, corresponding to the iteration numbers 1,2, 4,
and 6. We find that ${\cal A}_1$ is valid for $0.25 \le f \le 1.0$; ${\cal A}_2$ is
valid for $0.095491502\cdots \le f \le 0.65450849\cdots$; ${\cal A}_4$ is valid for
$0.030153689\cdots \le f \le 0.88302222\cdots$; and ${\cal A}_6$ is valid for
$0.014529091\cdots \le f \le 0.94272801\cdots $. These results strongly indicate that by
using ${\cal A}_{2n}$ of ever larger $n$, an ever larger range of $f$ can be covered
which in the limit of $n\rightarrow\infty$ approaches the full range $0\le f\le 1$,
but a general proof has not yet been obtained.  The validity or non-validity of this
statement, and the properties of the odd members of this family, will be discussed in a
future work. All members of this family of algorithms are characterized by two
phase parameters, $\theta$ and $\phi$. These two parameters are individually adjusted
in order to make each member a sure-success algorithm. I find that (i) at least
for the members ${\cal A}_2$, ${\cal A}_4$, ${\cal A}_6$, but most-likely also for all
higher even members of the family, $\phi = 2 \theta$ is an acceptable choice for $\phi$,
(ii) for each of the even members ${\cal A}_2$, ${\cal A}_4$, ${\cal A}_6$, and
most-likely also for each of all higher even members of the family, the required value
for $\theta$ for it to work is a unique function of $f$ just inside the boundary of its
validity range of $f$, but the number of acceptable values of $\theta$ graduately
increases to $n$ deep inside the validity $f$-range for ${\cal A}_{2n}$. The
algorithm member ${\cal A}_1$, on the other hand, requires $\phi = -2 \theta$, then
$\theta$ depends uniquely on $f$ within the validity range. No other odd members have
yet been analyzed. In all cases studied, I find the required $\theta$ and $\phi$ to
be independent of $N$, and to only depend on $f$. There is strong indication that this
statement is true for all members of the family.

All members of this family $\{{\cal A}_n\,|\,n = 1,2,3,\cdots\}$ are achieved with two
unitary operators which generalize the two corresponding operators introduced by
Grover:

In a Hilbert space spanned by a set of $N$ orthonormal basis states
$\{\,|i>\,|\,n = 1, 2, 3, \cdots, N\}$, each of which represents one element of the
data base, Grover introduced an {\it unitary} operator, which I shall denote as
$\hat F_{\nu}$, which changes the sign of the $\nu$'s amplitude $C_{\nu}$ in any quantum
state $|\Psi> = \sum_{i=1}^N C_i\,|i>$. This operator is generalized to the operator
$\hat F_{\phi}^{(a)}$, which introduces the extra phase factor $-e^{i\phi}$ to each of
the amplitudes $\{C_{\nu}\,|\,\nu\in a\}$, where $a$ denotes the set of acceptable
elements in the data base.~\cite{note2}  Mathematically,
\begin{equation}
\label{1stuniop}
\hat F_{\phi}^{(a)} \equiv \hat I - (e^{i\phi} + 1)\sum_{\nu\in a}|\nu><\nu|\,.
\end{equation}
where $\hat I \equiv \sum_{i = 1}^N\,|i><i|$ is the identity operator. For $\phi = 0$,
and $a$ containing only one element $\nu$, this operator reduces to the operator
$F_{\nu}$ introduced by Grover.

A second {\it unitary} operator introduced by Grover is the ``inversion about the mean''
operator, which can be written in the form:
\begin{equation}
\label{invabtmn}
\hat O \equiv \sum_{i,j}[(2/N) - \delta_{i,j}]\,|i><j|.
\end{equation}
I generalize it to
\begin{equation}
\label{2nduniop}
\hat O_{\theta} \equiv \sum_{i,j}[(2\cos\theta/N) - e^{i\theta}\delta_{i,j}]\,|i><j|\,,
\end{equation}
which reduces to Grover's ``inversion about the mean'' operator if $\theta = 0$.
That $\hat O_{\theta}$ is unitary can be easily verified. It is also easy to show that
it is the most general unitary operator of the form
$\sum_{i,j}[(A + B\delta_{i,j}]\,|i><j|$, if one disregards an unimportant overall phase
factor. I am not aware of any earlier published work introducing this umitary operator.

Since $\hat F_{\phi}^{(a)}$ and $\hat O_{\theta}$ are both complex operators, I also
need their hermitian conjugate operators, $F_{\phi}^{(a)\,\dag}$ and
$\hat O_{\theta}^{\dag}$, which are also the inverse operators of $\hat F_{\phi}^{(a)}$
and $\hat O_{\theta}$, respectively. Actually they are simply $\hat F_{-\phi}^{(a)}$
and $\hat O_{-\theta}$.

Before any algorithm is applied, every element in the data base should be regarded as
to have equal probability of being the right choice.  Grover represented this fact by
starting with the quantum state:
\[
|\Psi_0> = (1/\sqrt{N})\sum_i^N |i>\,,
\]
i.e., the state with every $C_i = 1/\sqrt{N}$, so that the probability of finding any
element of the data base is $|C_i|^2 = 1/N$. The quantum algorithm he introduced is to
repreatedly apply the unitary operator product $\hat O\hat F_{\nu}$ $n$ times on the
state $|\Psi_0>$, followed by a measurement to cause the state to collapse to one of
the basis states. He showed that when $n$ is of an optimal value of the order of
$\sqrt{N}$, All $|C_i|^2$ will be very close to zero except the particular one
$|C_{\nu}|^2$, corresponding to the desirable element $\nu$ in Grover's search problem,
which will be very close to unity. However, except for some special values of $N$, one
will not obtain exact unity for $|C_{\nu}|^2$, and exact zero for all other $|C_i|^2$.
Thus Grover algorithm is in general not a sure-success alorithm, even in theory, when
potenial implementation errors are not taken into account.
We generalize Grover's algorithm to a family of {\it sure-success} algorithms, each
member of which is characterized by an integer $n$. Denoting these member algorithms as
$\{{\cal A}_n\}$, then the even [$(2n)$th] member $\{{\cal A}_{2n}\}$ are defined as
applying the unitary operator product
$\hat\Lambda\equiv\hat O_{\theta}^{\dag}\hat F_{\phi}^{(a)\,\dag} \hat O_{\theta}
 \hat F_{\phi}^{(a)}$
$n$ times to the state $|\Psi_0>$, followed by the same measurement used in the Grover
algorithm. The odd [$(2n+1)th$] member $\{{\cal A}_{2n+1}\}$, is to
apply the unitary operator product $\hat O_{\theta}\hat F_{\phi}^{(a)}\hat\Lambda^n$
to the state $|\Psi_0>$, before the same measurement is made. Thus $\{{\cal A}_{n}\}$
makes $n$ queries of the data base. ``Sure success'' of each of these algorithms is
achieved by adjusting the two parameters $\theta$ and $\phi$ so that all $|C_i|^2$,
with $i$ not belonging to the set $a$ of the generalized Grover search problem
introduced here, are exactly zero. All $|C_i|^2$ with $i\in a$ will then be exactly equal
to $1/(fN)$, where $fN\equiv N_a$ is the number of elements in the set $a$, since
probability is conserved by unitary operations. Below we
show how this is done explicitly for the four members ${\cal A}_1$, ${\cal A}_2$,
${\cal A}_4$, and ${\cal A}_6$. After that I will speculate about all even members
${\cal A}_{2n}$ of the family, leaving the odd members higher than the first to be
discussed in a later work.

\begin{figure}[htb]
\centerline{\epsfig{figure=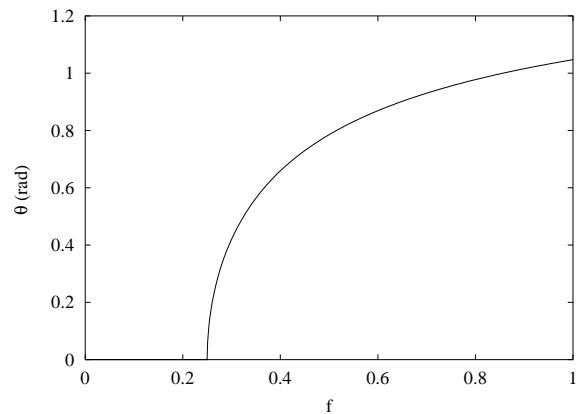, angle=270, width=8cm}}
\vspace{0.2cm}
\caption{Plotted is $\theta$ versus $f$ for the algorithm ${\cal A}_1$.}
\label{FIG:theta1}
\end{figure}

Consider first the algorithm member ${\cal A}_1$. One has the identity:
\begin{equation}
\label{EQ:OF1}
\hat O_{\theta}\hat F_{\phi}^{(a)}|\Phi_0> = [2\cos\theta(1-f-fe^{i\phi}) -
e^{i\theta}\hat F_{\phi}^{(a)}]|\Phi_0>\,.
\end{equation}
Since the operator $\hat F_{\phi}^{(a)}$ is equivalent to an identity operator in the
subspace corresponding to all unacceptable elements of the data base, sure success
of this algorithm is achieved by demanding
\begin{equation}
\label{EQ:cond1}
2\cos\theta(1-f-fe^{i\phi}) - e^{i\theta} = 0\,.
\end{equation}
which has the solution $\phi = -2 \theta$, and
\begin{equation}
\label{EQ:theta1}
\theta = (1/2)\cos^{-1}[(1/2f) - 1]\,.
\end{equation}
Note that if ($\phi$, $\theta$) is a solution, then ($-\phi$, $-\theta$) is also a
solution. This is true for all higher members of the family also, and one can easily
see why.
Equation~(\ref{EQ:theta1}) has solution only for $1/4\le f\le 1$, which is the validity
range of this algorithm.
Within this range, I have plotted $\theta$ as a function of $f$ in Fig.~\ref{FIG:theta1}
assuming $\theta > 0$.
The following special cases are of interest: (i) For $f=1/4$, I find $\phi=\theta=0$,
and the operators reduce to those introduced by Grover, and this algorithm becomes a
special case of Grover's algorithm. (ii) For $f=1/3$, I find $\phi = \pm\pi/3$ and
$\theta = \mp\pi/6$. (iii) For $f = 1/2$, I find $\phi = \pm\pi/2$, and
$\theta = \mp\pi/4$. (iv) For $f = 2/3$, I find $\phi = \pm 104.477\cdots^{\circ} =
\pm 0.580430...\pi$ and $\theta = \mp 52.2387\cdots^{\circ} = \mp 0.290215\cdots\pi$.
Finally, (v) for $f = 1$, I find $\phi = \pm 2\pi/3$, and $\theta = \mp \pi/3$, but in
this case the operator product $\hat O_{\theta}\hat F_{\phi}^{(a)}$ acting on $|\Phi_0>$
simply reproduces $|\Phi_0>$.

Next, let us consider the second member ${\cal A}_2$. One has the identity:
\begin{eqnarray}
\label{EQ:OF2}
\hat\Lambda|\Phi_0> &=& \{[(2\cos\theta)^2|(1-f-fe^{i\phi})|^2 - e^{2i\theta}]\nonumber\\
&& - (2\cos\theta)e^{-i\theta}(1-f-fe^{i\phi})\hat F_{\phi}^{(a)\dag}\}|\Phi_0>\nonumber\\
&\equiv& (A_1 - B_1\hat F_{\phi}^{(a)\dag})|\Phi_0>\,.
\end{eqnarray}
(Note that $A_1 = |B_1|^2 - e^{2 i \theta}$.)
Thus to ensure that this is a sure-success algorithm, one needs only demand
$A_1 - B_1 = 0$. The imaginary part of this condition can be written as
\begin{equation}
\label{EQ:Im-cond2}
{\rm Im}(A_1 - B_1) = (2 f \cos\theta)[\sin(\phi - \theta) - \sin\theta)] = 0 \,.
\end{equation}
so it can be satisfied with $\phi = 2\theta$. (It is easy to see that $\cos\theta\ne 0$.)
Then the real part of this condition reduces to
\begin{equation}
\label{EQ:Re-cond2}
{\rm Re}(A_1 - B_1) = 1 + 4 f \mu^2 - 16 f (1 - f) \mu^4 = 0
\end{equation}
where $\mu\equiv \cos\theta$. It has the solution
\begin{equation}
\label{EQ:theta2}
\theta = \frac{1}{2}\cos^{-1}\{\frac{1}{4(1-f)}[\sqrt{\frac{4}{f}-3}+(4f-3)]\}\,,
\end{equation}

\begin{figure}[htb]
\centerline{\epsfig{figure=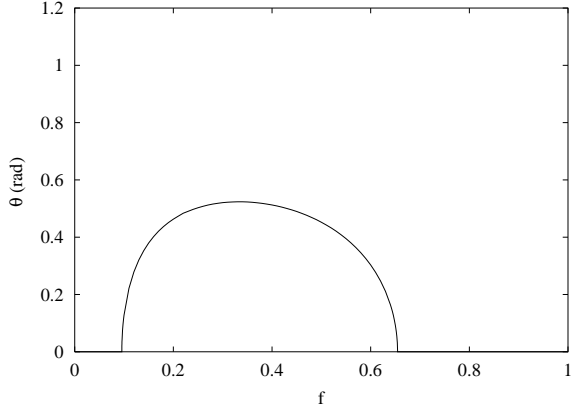, angle=270, width=8cm}}
\vspace{0.2cm}
\caption{Plotted is $\theta$ versus $f$ for the algorithm ${\cal A}_2$.}
\label{FIG:theta2}
\end{figure}

This equation has solution only if $0.095491502\cdots\le f \le 0.65450849\cdots$. Within
this range, I have plotted $\theta$ as a function of $f$ for this algorithm in
Fig.~\ref{FIG:theta2}, assuming $\theta > 0$.
%

%
%
Note that with the lagorithm ${\cal A}_2$ we can cover $f$ down to slightly below $0.1$.
 
\begin{figure}[htb]
\centerline{\epsfig{figure=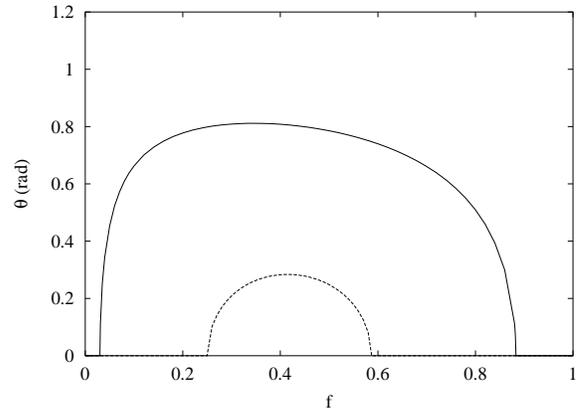, angle=270, width=8cm}}
\vspace{0.2cm}
\caption{Plotted is $\theta$ versus $f$ for the algorithm ${\cal A}_4$.}
\label{FIG:theta4}
\end{figure}

Next, let us consider the algorithm member ${\cal A}_4$, leaving ${\cal A}_3$ and higher
odd members for future discussion, since they are deemed less important. I have first
established the following theorem: If
$\hat\Lambda ^n |\Psi_0> = [A_n - B_nF_{\phi}^{(a)\dag}]|\Psi_0>$, then
\begin{eqnarray}
\hat\Lambda ^{n+1} |\Psi_0> &=& \{[A_1A_n - e^{-2i\theta}B_1^*B_n]\nonumber\\
  && - [B_1A_n - e^{-2i\theta}B_n] F_{\phi}^{(a)\dag}\}\,|\Psi_0>.
\label{lambda_n+1}
\end{eqnarray}
That is,
\begin{equation}
\pmatrix{A_{n+1}\cr B_{n+1}\cr} =
        \pmatrix{A_1& -B_1^*e^{-2i\theta}\cr
                 B_1& -e^{-2i\theta     }\cr} \pmatrix{A_n\cr B_n\cr}\,.
\label{EQ:AB-matrix}
\end{equation}
Thus $A_2 = |B_1|^4 - [2\cos(2\theta) + e^{2 i \theta}]|B_1|^2 + e^{4 i \theta}$ and
$B_2 = [|B_1|^2 - 2 \cos(2\theta)]B_1$. To ensure sure-success for this algorithm, one
needs to require $A_2 - B_2 = 0$. It is easy to show that
\begin{equation}
\label{EQ:Im-cond4}
{\rm Im}(A_2 - B_2) = [|B_1|^2 - 2\cos(2\theta)]{\rm Im}(A_1 - B_1)\,.
\end{equation}
I shall consider in a future work the possibility of satisfying this equation by
setting the first factor equal to zero.  Here I concentrate on the fact that due
to its second factor this equation can be satisfied by letting $\phi = 2 \theta$.
Then $\theta$ is given by

\begin{eqnarray}
\label{EQ:Re-cond4}
&& {\rm Re}(A_2 - B_2) = 1 + 8 f \mu^2 - 48 f (1 - f) \mu^4\nonumber\\
&& \ \ - 64 f^2 (1 - f) \mu^6 +
                256 f^2 (1 - f)^2 \mu^8 = 0\,.
\end{eqnarray}
I have plotted $\theta$ as a function of $f$ for this algorithm in Fig.~\ref{FIG:theta4}
assuming $\theta > 0$.
It is seen that solution exists only for
$0.030153689\cdots\le f \le 0.88302222\cdots$, and that in the narrower range
$0.25\le f \le 0.58682408\cdots$ A second solution for $\theta$ appears
for each $f$. It should be obvious that this algorithm is valid for those values
of $f$ only, for which at least one solution for $\theta$ exists, thus the larger $f$
range is also the validity range of this algorithm.

\begin{figure}[htb]
\centerline{\epsfig{figure=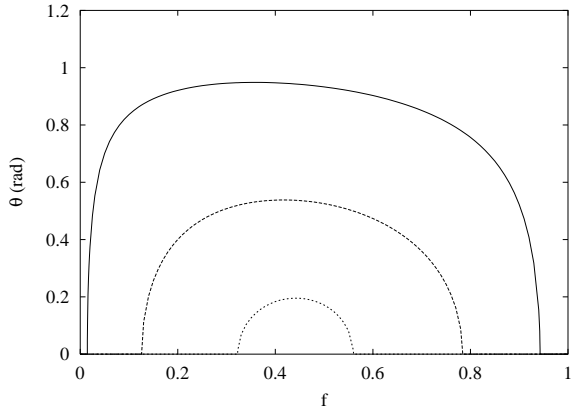, angle=270, width=8cm}}
\vspace{0.2cm}
\caption{Plotted is $\theta$ versus $f$ for the algorithm ${\cal A}_6$.}
\label{FIG:theta6}
\end{figure}

Finally, let us consider the algorithm member ${\cal A}_6$. Eq.~(\ref{EQ:AB-matrix})
allows me to obtain $A_3 = |B_1|^6 - [4\cos(2\theta) + e^{2 i \theta}] |B_1|^4 +
2[\cos(4\theta) + 1 + e^{4 i \theta}] |B_1|^2 - e^{6 i \theta}$, and
$B_3 = \{|B_1|^4 - 4\cos(2\theta)|B_1|^2 + [2\cos(4\theta) + 1]\}B_1$.
Thus I find
\begin{equation}
\label{EQ:Im-cond6}
{\rm Im}(A_3 - B_3) = \{[|B_1|^2 - 2\cos(2\theta)]^2 - 1\}{\rm Im}(A_1 - B_1)\,.
\end{equation}
Again, I shall not consider here letting the first factor equal to zero. Then again
$\phi = 2\theta$ from ${\rm Im}(A_3 - B_3)=0$, and $\theta$ is given by
\begin{eqnarray}
\label{EQ:Re-cond6}
&&{\rm Re}(A_3 - B_3) = 1 + 12 f \mu^2 - 96 f (1 - f) \mu^4 \nonumber\\
&&\ \ - 256 f^2 (1 - f) \mu^6 + 1280 f^2 (1 - f)^2 \mu^8 \nonumber\\
&&\ \ \ \ + 1024 f^3 (1 - f)^2 \mu^{10} - 4096 f^3 (1 - f)^3 \mu^{12} = 0\,.
\end{eqnarray}
I have plotted $\theta$ as a function of $f$ for this algorithm in Fig.~\ref{FIG:theta6}
assuming $\theta > 0$.
It is seen that solution exists only for
$0.014529091\cdots\le f \le 0.94272801\cdots$, which is the validity range of this
algorithm. In the narrower range $0.12574462\cdots\le f \le 0.78403237\cdots$
a second solution for $\theta$ appears for each $f$,
and in the even narrower range $0.32269755\cdots\le f \le 0.56026834\cdots$
a third solution for $\theta$ appears for each $f$.

A trend is clearly established by the above study of the first three even members.
It strongly suggests that for all even members, (i) $\phi = 2\theta$ is always a valid
solution, with $\theta$ depending on $f$, but not on $N$;
(ii) the $f$-range in which at least one $\theta$ value exists becomes ever larger if
${\cal A}_{2n}$ of ever larger $n$ is considered, with the $n\rightarrow \infty$ limit
being very likely the full range $0\le f \le 1$; (iii) in general the number of valid
choices for $\theta$ increases to $n$ deep inside the validity $f$-range for
${\cal A}_{2n}$. General proofs of these statements have not yet been obtained.

In summary, an infinite family of sure-success quantum algorithms is introduced here for
solving the generalized Grover search problem of finding any one element of a set of
acceptable choices which constitute a fraction $f$ of all elements in an unsorted data
base. This is achieved by two unitary operators each containing a phase parameter. These
operators are generalizations of the two operators introduced by Grover for his original
search problem. The two phase parameters are adjusted for each member of the family to
ensure its sure-success, which is found possible only within a different $f$-range for
each member of the algorithm family. An infinite sub-family (the ``even'' members)
appears to have the property that the validity $f$-range of a lower member is totally
embedded inside that of a higher member, with the limit being very likely the full range
$0\le f \le 1$. As long as $f$ is within the validity range, the lowest member of the
sub-family is then the most convenient, since it requires the least number of queries
of the data base.



\end{document}